\documentclass[12pt]{article}

\usepackage{amssymb,epsfig,cite}

\newcommand{\CC}{{\kern+.25em\sf{C}\kern-.45em\sf{{\small{I}}} \kern+.45em\kern-.25em}}

\def\lsi{\raise0.3ex\hbox{$<$\kern-0.75em\raise-1.1ex\hbox{$\sim$}}}
\def\gsi{\raise0.3ex\hbox{$>$\kern-0.75em\raise-1.1ex\hbox{$\sim$}}}
\def\backder{\raise1.4ex\hbox{$\leftarrow$\kern-0.75em\raise-1.4ex\hbox{$\partial$}}}
\def\forder{\raise1.4ex\hbox{$\rightarrow$\kern-0.8em\raise-1.4ex\hbox{$\partial$}}}

\newcommand{\C}{{\kern+.25em\sf{C}\kern-.45em\sf{{\small{I}}} \kern+.45em\kern-.25em}}
\newcommand{\R}{{\kern+.25em\sf{R}\kern-.78em\sf{I} \kern+.78em\kern-.25em}}
\newcommand{\NN}{{\kern+.25em\sf{N}\kern-.78em\sf{I} \kern+.78em\kern-.25em}}

\newcommand{\be}{\begin{equation}}
\newcommand{\ee}{\end{equation}}
\newcommand{\bea}{\begin{eqnarray}}
\newcommand{\eea}{\end{eqnarray}}

\begin{document}

\begin{center}
  
{\Large{\bf What are Elementary Particles?}} \\

\vspace*{4mm}

{\large{\bf From Dark Energy to Quantum Field Excitations}} \\

\vspace*{7mm}

Wolfgang Bietenholz \\
Instituto de Ciencias Nucleares \\
Universidad Nacional Aut\'{o}noma de M\'{e}xico \\
A.P.\ 70-543, C.P.\ 04510 Ciudad de M\'{e}xico, Mexico
\vspace*{2mm}

\end{center}

\noindent
{\em We describe the very nature of the elementary particles, which
  our (visible) Universe consists of. We point out that they are
  not point-like, and we depict their ways of interacting. We also
  address puzzles that occur even in the absence of particles,
  in the vacuum.} \\

\noindent
This article is meant to be qualitative and very simple; slightly
technical remarks are added as footnotes and as an appendix.

\section{Basic building blocks of matter}

If we break up any kind of matter into smaller and smaller pieces,
we ultimately reach a point of basic building blocks, which are
not divisible anymore: Democritus would have called them ``atoms'',
but for us these are the {\em elementary particles}.
So far 25 types of elementary particles have been experimentally
confirmed; the entire visible Universe consists of them.\footnote{This
  includes quarks and gluons, although they cannot be directly detected,
  as well as leptons, electroweak gauge bosons and the Higgs particle.
  We refer to the {\em visible} part of the Universe in order
  to exclude Dark Matter and Dark Energy; the latter will be
  addressed in Appendix A. Gravitation belongs to our daily
  experience, but the particle, which is held responsible for it ---
  the graviton --- has not been observed.}
They are incorporated in the Standard Model of Particle Physics;
prominent examples are the {\em electron} and the
{\em photon} (the particle of light).
As if this wasn't enough, the literature of theoretical
physics is replete with speculations about additional types of
elementary particles.

\begin{figure}[h!]
\begin{center}
 \includegraphics[angle=0,width=.45\linewidth]{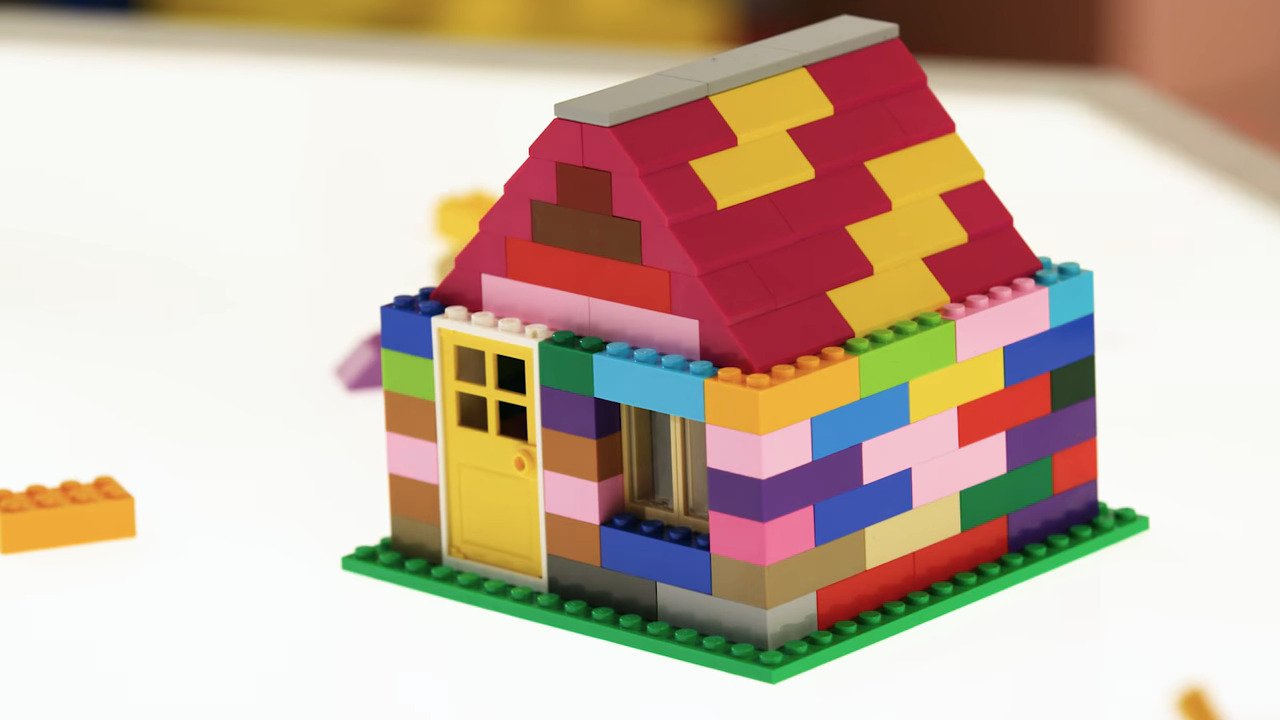}
\vspace*{-3mm}
 \end{center}
\caption{We can easily decompose a Lego house into its building blocks.
  If we keep on decomposing matter down to its most fundamental
  building blocks, we end up with elementary particles.}
\label{legofig}
\end{figure}

We do not go through the list of all these known elementary
particles (let alone the hypothetical ones); their table can
easily be found in many places. Instead
we want to address a question, which is seldomly discussed in
popular science: {\em what kind of objects are elementary particles?}
Amazingly, even in the physics literature this issue is treated
as an orphan: there are numerous textbooks devoted to particle physics,
which hardly clarify what these objects actually are.\footnote{The text
  is written in terms of ``particles'', and the formulae in terms of
  ``fields'', but the question how these terms are related is by
  no means as clear as it is supposed to be.}

The common intuitive picture, which is based on our perception
of macroscopic objects, views them as ``tiny balls''.
We are going to point out that this picture is erroneous,
and that they are not ``point-like objects'' either. The latter
(mysterious) claim is wide-spread, but that doesn't make it correct.

\section{Quantum Field Theory}

The mathematical formalism that successfully describes elementary
particles is called {\em Quantum Field Theory.} In the course of the
20th century it has replaced Quantum Mechanics.\footnote{In contrast to
  Quantum Mechanics and classical physics, Quantum Field Theory
  successfully incorporates the concepts of both quantum physics and
  Special Relativity. (A complete unification of quantum physics with
  General Relativity has not been achieved.)\label{fn:SpecRel}}

\begin{figure}[h!]
\begin{center}
 \includegraphics[angle=0,width=.5\linewidth]{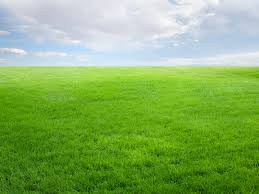}
 \end{center}
 \caption{A field in the usual sense. In physical field theories,
   the meadow-land is converted into space, and the grass blades
   into abstract mathematical variables, which we denote as
   ``oscillators''.}
\label{fieldfig}
\end{figure}

In order to symbolically interpret the term {\em field theory}, we could
view the entire Universe as the ``meadow-land'', endowed with some kind of
``grass blades'' everywhere. 
The latter take an abstract mathematical
form: certain variables are permanently attached to each point in space.
A ``field'' is one type of such variables.
Each variable, at a given point, can change its value as a function of time,
we could say that it ``vibrates'' or ``oscillates''. In the following
we are going to refer to an ``oscillator'', a term which can be reasonably
well justified, see {\it e.g.}\ Ref.\ \cite{Peskin}.

It is always risky to invoke a simplified picture for illustration purpose,
but we do so nevertheless. There is a rough analogy with sound in the air:
let us assume the absence of wind in some volume,
so the molecules of the air have (essentially) static equilibrium positions,
but their vibrations around them represent sound.
This bears some similarity with field theory, which we can further
strengthen by referring to sound in a crystalline solid, where ions
oscillate around their grid sites, with displacements analogous to
a field variable.
We repeat, however, that actual field variables are abstract
mathematical quantities.

\begin{figure}
\begin{center}
\vspace*{-7mm}
 \includegraphics[angle=0,width=.45\linewidth]{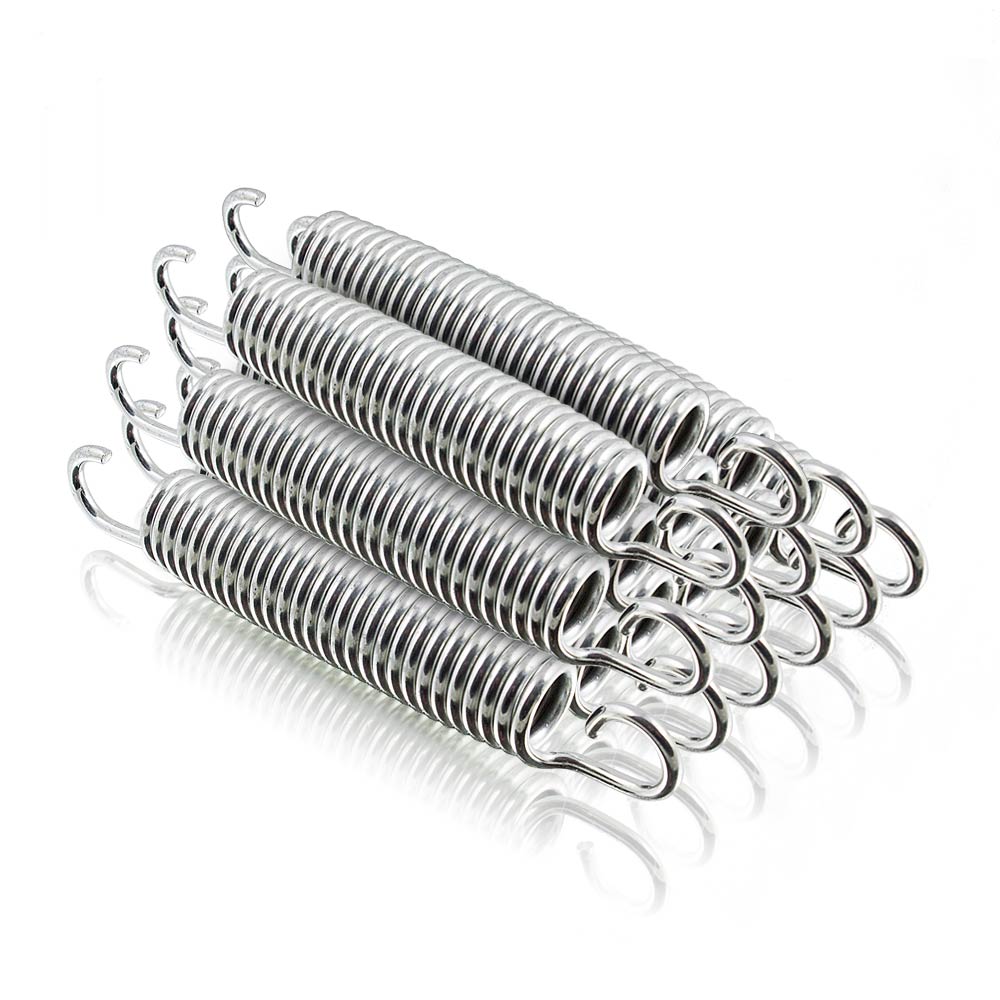}
 \end{center}
\vspace*{-1.2cm}
\caption{A set of classical oscillators; quantum oscillators are hard
  to depict.}
\label{oscillatorfig}
\end{figure}

So let us denote a field variable at one point as an {\em oscillator}.
It can be in its {\em ground state,} where its energy $E_0$ is minimal
(in quantum physics we have $E_0 > 0$),\footnote{We are referring
  to the simple case of bosonic fields, like the photon field
  or the Higgs field. There is another class of particles called
  {\em fermions}, which includes the electron, and which emerge
  from fields with $E_0 < 0$, see Appendix A.\label{fermifn}}
or in an {\em excited state} with a higher energy $E>E_0$. As we mentioned
before, the state of any of these oscillators (which fill the entire
space) is time-dependent.

\section{Vacuum and particles}

At this point, we only consider one field, {\it i.e.}\ one type
of oscillator.

Let us assume all the oscillators in some volume to be in their ground state.
We denote this as the {\em vacuum}, which means that the field takes
its state of minimal energy in this volume.\footnote{For simplicity
  we refer to a free ``neutral scalar field'', where the field is
  real valued. Other fields involve different types of
  variables, and when the fields are self-coupled or coupled to other
  fields, {\it i.e.}\ not free, then even the vacuum often takes a
  complicated structure.}
We would say ``nothing is there'', although the oscillators are
actually there, but none of them vibrates with any excitation energy
$E>E_0$. It is tempting to interpret the ground state energy throughout
the Universe as {\em Dark Energy;} this leads, however, to a
dreadful puzzle, which we address in Appendix A.

Now let us insert a single particle into this volume, say a particle at
rest (with respect to the volume). This requires a massive particle, like
the electron, and we denote its mass as $m > 0$.\footnote{Mathematically
  this is achieved by applying a so-called creation operator to the vacuum
  state. It 
  cannot be restricted to one spatial point.\label{fn:create}}

What does this mean for the field under consideration? It will be excited,
such that its total energy inside this volume takes its minimal value above
the vacuum energy. In quantum physics, this minimal excitation corresponds
to a finite energy gap $\Delta E$; the energy cannot be increased
continuously above the vacuum energy. We also know that this energy gap,
{\it i.e.}\ the particle's rest energy, is related to the particle mass
by a famous formula, $\Delta E = mc^2$ (where $c$ is the speed of
light in vacuum).

If these oscillators were all independent, the obvious way to
arrange for a minimal excitation would be to excite just one of
them to the first energy level and leave all the rest in their
ground state. However, this is not how it works: the oscillators
are {\em coupled} to the their nearby neighbours, hence exciting
one of them inevitably affects its vicinity
(cf.\ footnote \ref{fn:create}).\footnote{In mathematical
  terms, there are field derivatives contributing to the energy,
  hence a discontinuous spike --- or even just a very sharp peak ---
  is not suitable for an excitation with minimal energy.}

Instead we obtain a smooth excitation energy profile, which we assume
to have a maximum in its centre. It turns out that it decays exponentially
with the distance from this centre, where (in the free case)
the range of the decay is proportional to
the inverse particle mass, ${\rm range} \propto 1/m$.
This range coincides with the Compton wavelength \cite{CUP}.

\begin{figure}[h!]
\vspace*{-7.5cm}
\begin{center}
\hspace*{1.6cm}
  \includegraphics[angle=0,width=1.45\linewidth]{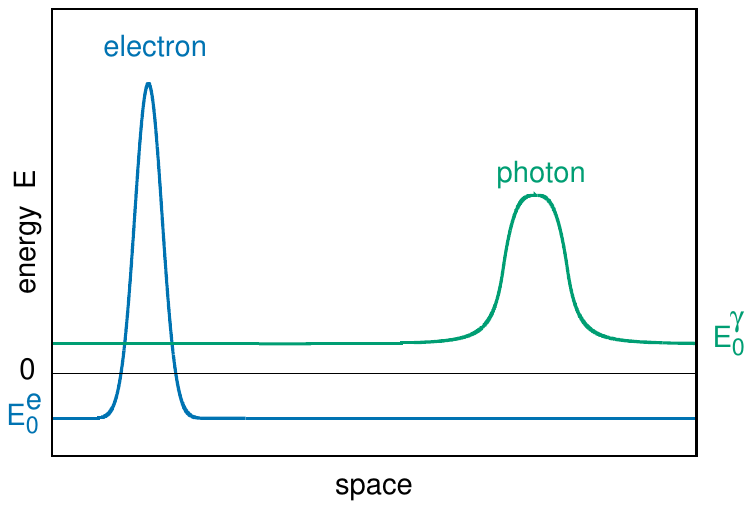}
\vspace*{-2.3cm}
\end{center}
\caption{A symbolic illustration of the energy profiles of an
  electron and a photon, as examples of a massive and a massless
  elementary particle ($E_{0}^{\rm e}$ and $E_{0}^{\gamma}$ are the
  corresponding ground state energies).}
\label{profiles}
\end{figure}

For a massless particle, like the photon, this decay is slower:
here it only follows some negative power of the distance from the
particle centre, but not an exponential decay \cite{CUP}.
In either case, with $m=0$ or $m>0$, we see that particles do have
an {\em extent,} they are {\em not} point-like objects.\footnote{This is
  correctly emphasised {\it e.g.}\ in Refs.\ \cite{ScharfSteinmann}.}
Such profiles are symbolically illustrated in Figure \ref{profiles}.

In the framework of this qualitative note, we stay with the
observation that an elementary particles is not a point particle,
only marginally touching upon the somewhat distracting question what
its size amounts to. We could, for instance, define the size of a
free electron with the exponential profile decay that we just
mentioned, or we could refer to the so-called {\em form factor,}
which is observed in electron scattering experiments.
However, even within this phenomenological approach, the ``electron size''
is still ambiguous: one would have to distinguish an ``electric''
and a ``magnetic radius'', depending on the scattering effect that
one refers to.\footnote{Ref.\ \cite{Weinberg} explains how to
  perturbatively compute such radii in Quantum Electrodynamics.}

Finally one might be tempted to refer to the resolution of a particle
detector, for instance to the size of a pixel in a raster image.
However, no matter how small the pixels are, a single photon will
always be detected in only one of them, so we cannot determine a
photon size in this way. This does not imply that the photon has
zero extent: its profile ``collapses'' into just one pixel at the
moment of the detection.\footnote{We don't know how exactly this
  happens, but it does happen. It is analogous to the notorious
  ``collapse of a wave function'' in Quantum Mechanics.}

\section{Particles in motion}

We have seen that elementary particles can be understood as small
regions, or zones,
where a (quantum) field is excited. These zones, {\it i.e.}\ the
particles, can {\em move} (say, relatively to each other). 
A descriptive picture is that the excited oscillators lose energy,
and eventually drop down to the ground state, transferring their
energy to nearby other oscillators, and so on. In this way, the
particle centre moves, and with it all the excitation zone.

This picture is reminiscent of an ordinary (though Lorentz invariant)
wave,\footnote{Frank Wilczek occasionally denotes a quantised field
  excitation as a {\em ``wavicle''} \cite{Wilczek}, which makes sense,
  but the established term is ``particle''.}
but it is important to stress:
if the terms ``particle'' and ``wave'' are understood from a
classical (not quantum) perspective, {\it i.e.}\ as concepts, which
match our macroscopic perception, then neither of these two terms
describes a quantum particle appropriately. For the lack of
quantum terminology in colloquial language, we are using those
terms, which have led to never-ending confusion.

\section{Multiple fields and particle interactions}

We now go beyond the consideration of just one particle type.
From Section~1 we know that there are at least 25 different fields,
{\it i.e.}\ at least 25 types of ``oscillators'' being present at
each point of the Universe, at any time. Some of these fields are
self-coupled or
{\em coupled to each other,} this is specified in the Standard Model.
Each field has its zones of excitations, corresponding to
a huge number particles in the Universe,\footnote{Even the ``empty''
  cosmic space is packed with about 411 photons and 366
  neutrinos per cm$^3$, see {\it e.g.}\ Ref.\ \cite{WBnu}.
  These are the most abundant types of particles.}
which may all be in motion.

If such excitation zones of coupled fields {\em overlap}, particles can
directly {\em interact} and we talk about {\em particle collision.}
The rule that instantaneous interactions do not happen over
a distance is known as {\em locality}. When a collision takes
place, energy can be transferred from one field to another. This
dynamics can also affect fields, that have not been excited before in
this zone, so additional types of particles can be
generated.\footnote{Capturing the dynamics of particle creation and
  annihilation is an essential achievement of Quantum Field Theory,
  in contrast to traditional Quantum Mechanics. This property is
  intimately related to the statement in footnote \ref{fn:SpecRel},
  see {\it e.g.}\ Ref.\ \cite{Nachtmann}.}

Obviously the creation of a heavy particle requires a lot of energy.
Therefore some laboratories, like CERN, accelerate massive particles
almost to the speed of light and arrange for collisions at extremely
high energy, in order to investigate whether this creates heavy
particles, which have not yet been observed --- possibly one of the
hypothetical particles that theoretical physicists speculate about.
The famous Higgs particle, which had been predicted since 1964
\cite{Higgs}, was finally observed in this manner at CERN in 2011/12
\cite{HiggsCERN} (a popular science description is given in
Ref.\ \cite{HiggsWB}).

Once a rather heavy particle is generated, it tends to decay very
rapidly (unless there is a conservation law preventing this).
Then it transfers its energy to other fields, thus
creating several lighter particles, which corresponds to a
process of energy diffusion.\footnote{For further reading about
particle scattering and decay, we recommend Ref.\ \cite{Maggiore}.}
For instance the Higgs particle, with a mass of $125~{\rm GeV}/c^2$
--- the second-most heavy elementary particle that we
know\footnote{$1\, {\rm GeV}=10^{9} \, {\rm eV}$ (electronvolt)
  is a unit of energy,
  $1 \, {\rm eV} \simeq 1.6 \cdot 10^{-19} \, {\rm J}$.
  For comparison, the electron and the proton have a masses of
  $0.000511~{\rm GeV}/c^{2}$ and $0.938~{\rm GeV}/c^{2}$, respectively.}
--- has a lifetime is only about $10^{-22}~{\rm sec}$.

\section{The interwoven Universe}

Another manifestation of particle interactions are attractive
or repulsive {\em forces.}\footnote{The question how forces
  really emerge in Quantum Field Theory is another subject,
  which is not well covered in the literature, despite its
  importance. A sound pedagogical explanation is given in
  Ref.\ \cite{Zee}.} In contrast to Newton's
formulation of gravity, such forces do not act instantaneously
at a distance --- field theory is consistent with the principle
of locality, as we pointed out before.
For instance, the Coulomb force between two electrons is
indirect: each electron affects at its location the photon field
(they are coupled by the electric charge of the electron).
When the electrons move closer, the energy of the coupled
field system is enhanced, which implies a repulsive force\footnote{As
  a simple classical picture:
  if two static electrons are next to each other, an electric field
  $\vec E(\vec x)$ emerges, which is almost doubled compared to a
  single electron. Thus the field energy
  $\propto \ \int d^3x \, \vec E(\vec x)^{2}$ decreases when the
  electrons move apart.} (in jargon, this is due to the
``exchange of virtual photons'').

We know that the electromagnetic force can also be attractive,
for instance between the electron and its anti-particle, the
positron, which carries positive electric charge.\footnote{When
  they are close, they form a small, electrically neutral compound,
  which hardly affects the photon field, hence their proximity
  is energetically favoured.}
Other types of intermediate fields (so-called ``gluon fields'')
give rise to the {\em ``strong interaction''}, and in particular to
strong attraction, which (in suitable circumstances)
outweighs electromagnetic repulsion. Due to such forces,
some elementary particles form {\em bound states,} which are composite
particles. The best known examples are the proton and the neutron,
which build the atomic nuclei. Together with the electrons we
obtain atoms, which can be further clustered to molecules.
Following this sequence of composition, we reconstruct the larger
structures of matter, which we have decomposed in the very beginning
of this article.

However, from a fundamental perspective, such composite objects,
and the entire visible Universe, ultimately consist of the elementary
particles that we have described before. This is the particle
physicist's view of the world: the interactions among these
particles imply a very complicated, interwoven dynamics,
following probabilistic rules, which we investigate.

At last, returning to the simplistic analogy with sound,
we could --- figuratively speaking --- call this interwoven
dynamics the ``cosmic symphony'', or ``cosmic salsa
concert'', whatever you prefer.\footnote{Symbolically,
  this seems to bear some similarity with the concept
  of the ``harmony of spheres and numbers'' or ``musica
  universalis'', a philosophical concept which was supposedly
  advocated by the Pythagoreans over 2500 years ago \cite{vdWaerden}.
  Unlike them, however, we do not focus on the motion of celestial
  bodies, and we discard mystical interpretations.}

\noindent
    {\bf Acknowledgement:} I am indebted to Uwe-Jens Wiese for sharing
    his insight into the concept of ``wavicles''. I thank
    Ernesto Altshuler, Tamer Boz, Jaime Nieto-Castellanos and Lilian Prado
    for reading the manuscript, as well as the
    authors of Ref.\ \cite{Arkani} and the American Physical Society for
    the permission to reproduce Figure \ref{coincidencefig}.
    This work was supported by UNAM-DGAPA-PAPIIT, grant number IG100219.

\appendix

\section{The mystery of Dark Energy}

The consideration in the second paragraph in Section 3 suggests
the presence of a non-zero energy density $\rho_{\rm E_{0}}$ throughout
the Universe. Actually $\rho_{\rm E_{0}}$ even seems to diverge:
for a given field, a Fourier transform leads to a set of oscillators
with all possible frequencies.

For the free, neutral, massless scalar field, one oscillator contributes
the ground state energy $E_{0} = \frac{1}{2} \hbar \omega$, where
$\omega = \sqrt{\omega_{1}^{2}+\omega_{2}^{2}+\omega_{3}^{2}}$,
$\omega_{i}$ being the angular frequencies in different directions
(in 3 spatial dimensions),
and $\hbar$ is Planck's constant. If we compute $\rho_{\rm E_{0}}$
for the photon field by integrating $\int d^{3}\omega \ \hbar \omega$,
we obviously run into a divergence (a factor 2 accounts for
the two photon polarisation states).

It seems natural to introduce an {\em ultraviolet cutoff,} say at
the Planck scale $E_{\rm Planck} = \sqrt{\hbar c^{5} / G}
\simeq 1.2 \cdot 10^{28} \, {\rm eV}$, where $G$ is Newton's gravitational
constant. This restricts the
integral to $4 \pi \int_{0}^{E_{\rm Planck}} d \omega \, \hbar \omega^{3}$.
Taking into account the Fourier normalisation factor $1/(2\pi )^3$, we
obtain --- due to the ground state energy of the free photon field ---
the energy density
\begin{displaymath}
\rho_{\rm E_{0}} \approx \frac{E_{\rm Planck}^{4}}{8 \pi^{2} (\hbar c)^{3}} \ .
\end{displaymath}

In usual studies of quantum physics, such an additive constant in
the energy is irrelevant, since we are only concerned with energy
{\em differences}.\footnote{Considering well-defined differences
  of physical quantities, while putting aside, or isolating, a divergent
  additive constant, is the basic idea of {\em renormalisation}.}
If we add a constant term to the potential of
some system, then this does neither affect the (field) equations
of motion, nor the expectation values in Quantum Field Theory.
This changes, however, when we include {\em gravity:} note that a constant
energy density $\rho$ cannot be added to the potential in the simple
space-time integrated form $\propto \int dt \, d^3x \, \rho\,$ ---
that term is not covariant. Instead the space-time metric must be
involved, which is therefore affected by the quantity $\rho$
(in General Relativity even the metric is dynamical).

As a consequence, such a constant leads to a prominent
physical effect, namely (if it is sufficiently large)
the {\em accelerated expansion of the Universe.}
The driving energy density is denoted as {\em Dark Energy,} which can
be interpreted (up to a constant factor) as Einstein's
{\em Cosmological Constant.}\footnote{Albert Einstein
  introduced such a constant in his formulation of
  General Relativity, in order to construct a static Universe
  \cite{Einstein17}. Once Edwin Hubble and others convinced him
  that the Universe is rapidly expanding, and it turned out that his
  static solution would be unstable, he dismissed this constant and
  accepted the expanding solutions to his theory by Alexander Friedmann
  and Georges Lema\^{\i}tre \cite{Friedmann}
  (see Ref.\ \cite{Nuss} for a historic account).
  Thereafter, for almost seven decades, the Cosmological Constant was assumed
  to vanish, which implies a decelerated expansion. Einstein's original
  value was positive, just at the (unstable) transition between a
  decelerated and an accelerated expansion.
  Today a somewhat larger Cosmological Constant is appreciated as
  the most obvious explanation for the observed accelerated expansion.
  If we understand General Relativity as a low-energy effective theory,
  then the presence of this constant is natural.}
It corresponds to a negative pressure, which is occasionally denoted
as ``gravitational repulsion''.

So at the {\em qualitative} level, $\rho_{E_{0}}$ seems to provide a
neat explanation for this accelerated expansion, which was discovered at
the very end of the 20th century \cite{accUni}.
The 2011 Physics Nobel Prize was awarded for this observation.

However, our enthusiasm comes to an abrupt end when we proceed to the
{\em quantitative} level: the observation of Refs.\ \cite{accUni},
which is based on the distance and redshift of a set of
type Ia supernovea, corresponds to a vacuum energy density of about
$\rho_{\rm obs} \approx (0.002 \, {\rm eV})^{4}/(\hbar c)^{3}$.
Now we are stunned by a tremendous discrepancy from the
theoretical estimate $\rho_{\rm E_{0}}$,
\begin{displaymath}
\frac{\rho_{E_{0}}}{\rho_{\rm obs}} \approx 10^{121} \ .
\end{displaymath}
{\em This is perhaps the worst discrepancy between a theoretical prediction
  and an observed value in the history of science.}

$E_{\rm Planck}$ is the most natural energy cutoff, but a conceivable
alternative might be $E_{\rm GUT} \simeq 10^{25}\, {\rm eV}$, the energy where
the three gauge couplings of the Standard Model are predicted to converge
to the same strength. That reduces the above discrepancy to
$\rho_{E_{0},\, {\rm GUT}} / \rho_{\rm obs} \approx 10^{109}$, which is no
salvation. If we really wanted to maintain the previous derivation,
we had to lower the energy cutoff to $\approx 0.006 \, {\rm eV}$
instead of $E_{\rm Planck}$, but such a ridiculously low cutoff
does not make any sense: even the rest energy of an
electron is almost $10^{8}$ times higher.\footnote{At this point, the
  question arises whether also the field theoretic vacuum energy has been
  observed, since the Casimir force is now confirmed experimentally
  \cite{Lamoreaux}. Then this discrepancy would be even more puzzling.
  However, this conclusion is not compelling, since the Casimir effect
  can also be derived without referring to the vacuum energy of the
  photon field \cite{Casimirnovac}.}

\begin{figure}[h!]
  \vspace*{-1cm} 
\begin{center}
 \includegraphics[angle=0,width=.7\linewidth]{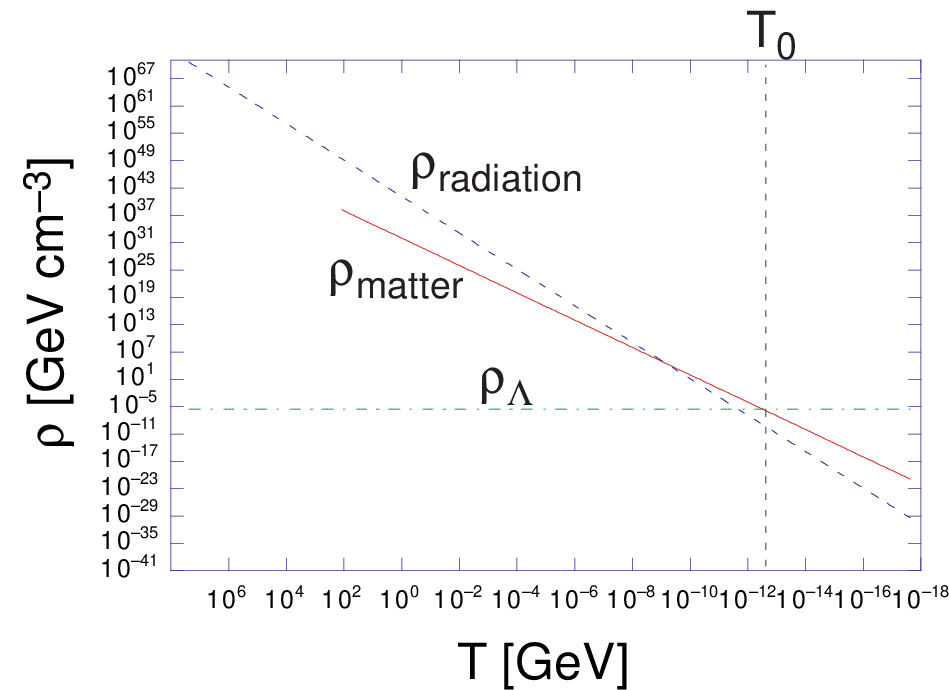}
  \vspace*{-5mm}
 \end{center}
\caption{Illustration of the coincidence problem, adapted from
  Ref.\ \cite{Arkani}. That work (and many others) distinguishes
  ``radiation'' (fast moving, {\it i.e.}\
  relativistic particles, mostly photons and neutrinos) from ``matter'':
  in the former (latter), most energy is kinetic (contained in the
  rest mass). In these terms, the first $\approx  47\,000$ years after
  the Big Bang were radiation-dominated, followed by the matter-dominated
  era, which lasted until $\approx 9.8 \cdot 10^{9}$ years after the
  Big Bang. Today, the age of the Universe is $\approx 13.8 \cdot 10^{9}$
  years, the cosmic microwave background has the temperature
  $T_{0} \simeq 2.7~{\rm K}$ (indicated in the figure), and our
  era is dominated by Dark Energy ($\rho_{\Lambda}$), to about 70\%.
  (For the Hubble constant, Ref.\ \cite{Arkani} inserted
  $H_{0} = 65~{\rm km/(s \ Mpc)}$.)}
\label{coincidencefig} 
\end{figure}

To make it even worse, there is in addition the {\em coincidence problem:}
by default, the Cosmological Constant, and therefore the Dark Energy density,
are assumed to be really {\em constant} during the evolution of the
Universe, whereas the matter density decreases due to its expansion;
it has decreased by many orders of magnitude since the time of the
Early Universe, see Figure \ref{coincidencefig}.
At that time matter density\footnote{Here we include radiation,
  unlike the terminology of Ref.\ \cite{Arkani}, but it doesn't matter
for the statements in this paragraph.} dominated over vacuum
energy, and in the far future it will be the other way round.
It so happens that just in our time the matter density
--- which is dominated to about 85\% by Dark Matter (which does
not interact with the photon field) ---
is of the same magnitude as the Dark Energy density
(they only differ by about a factor of 3). Is it by accident that
we just have the privilege to witness this transition, or does
this ``coincidence'' require an explanation?\footnote{Meanwhile
  a number of cosmologists speculate that the Cosmological ``Constant''
  might have changed in the course of the evolution of the Universe
  \cite{Sola} (this is reminiscent of the coupling ``constants'' in
  Quantum Field Theory, which actually depend on the energy scale;
  in jargon, they are ``running'').}
(Some people try to argue with the ``anthropic principle'', which
seems like an act of desperation.)

So far we have considered the example of a {\em bosonic} field, in
particular the photon. As we anticipated in footnote \ref{fermifn}, 
there are other types of particles called {\em fermions}
(the electron is an example), where such a huge vacuum energy
density emerges with a {\em negative} sign, so one
might hope for a large amount of cancellation.

If we were living in a perfectly {\em supersymmetric} world, the bosons and
fermions would appear in pairs with the same mass, and indeed the vacuum
energy would exactly cancel.
However, even if supersymmetry exists, it has be to be {\em broken}
in the low-energy regime where we are living:
for instance the bosonic partner of the electron,
the ``selectron'', must be much heavier than the electron --- if
it exists at all --- otherwise it would have been observed. The
extent of supersymmetry breaking, which is required to avoid
contradictions with observations, would allow for a strong reduction
of the ratio $\rho_{\rm E_{0}}/\rho_{\rm obs}$. It still has to be huge,
though, estimates suggest at least $\approx 10^{60}$ \cite{Carroll}
(even before knowing the LHC results),
so supersymmetry does not overcome this problem either. Also
the string community tried to solve the Cosmic Constant problem, without
arriving at any key clue \cite{Carroll,Witten}.
\begin{figure}[h!]
\begin{center}
 \includegraphics[angle=0,width=.83\linewidth]{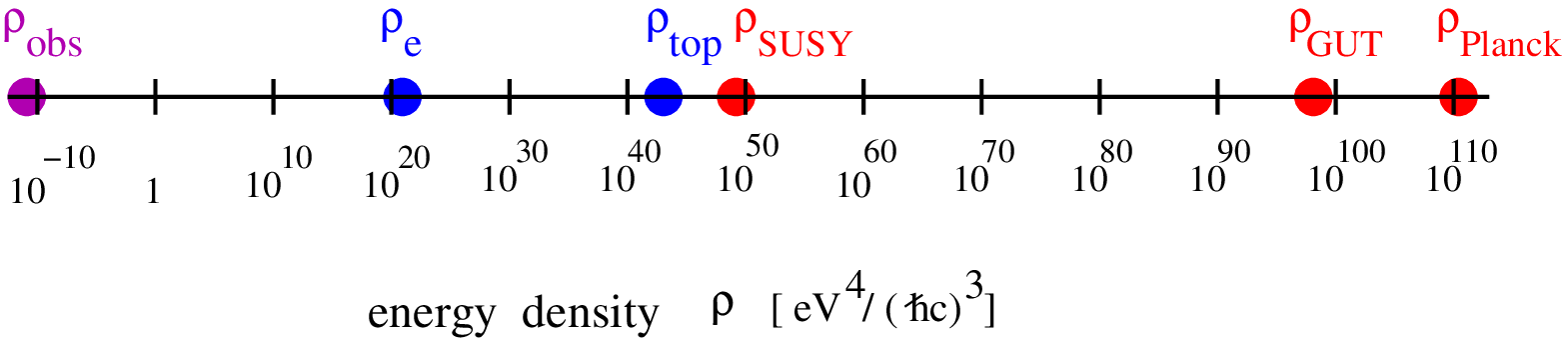}
  \vspace*{-5mm}
 \end{center}
\caption{The energy density as observed in the Universe, $\rho_{\rm obs}$,
  and as obtained from the vacuum energy in Quantum Field Theory, if we
  insert a cutoff at the rest energy of the electron, $\rho_{e}$,
  or of top-quark  (with a mass of 173~GeV/$c^{2}$
  the heaviest known elementary particle), $\rho_{\rm top}$.
  These cutoffs are not motivated, they are included just for
  comparison. Scenarios that could be considered as somehow motivated
  are referring to broken supersymmetry ($\rho_{\rm SUSY}$ corresponds to
  the lower bound for its breaking), a cutoff at the Grand Unification
  scale, $\rho_{\rm GUT}$, or at Planck scale, $\rho_{\rm Planck}$.
  In all these cases, the theoretical energy density
  exceeds $\rho_{\rm obs}$ by {\em many} orders of magnitude.}
\label{Escalefig}
\end{figure}

It is outrageous that we do not have any convincing solution to this
stunning puzzle, so this appendix finishes without a happy ending.
For reviews of this outstanding issue we refer to Refs.\
\cite{Sola,Carroll}.


\newpage

\begin{thebibliography}{11}

\bibitem{Peskin} M.\ E.\ Peskin and D.\ V.\ Schroeder,
  An Introduction to Quantum Field Theory,
  Westview Press, 1995 (Chapter 2).
  
\bibitem{CUP} W.\ Bietenholz and U.-J.\ Wiese,
  Quantum Field Theory and the Standard Model of Particle Physics:
From Fundamental Concepts to Dynamical Mechanisms,
to appear in Cambridge University Press (Chapter 4).

\bibitem{ScharfSteinmann} G.\ Scharf,
  Finite Quantum Electrodynamics: The Causal Approach, Springer,
  1995 (Chapter 3).\\
  O.\ Steinmann, Perturbative Quantum Electrodynamics
  and Axiomatic Field Theory, Springer, 2000 (Chapter 1). 

\bibitem{Weinberg} S.\ Weinberg, The Quantum Theory of Fields,
volume I, Cambridge University Press, 1995 (Chapter 11).

\bibitem{Wilczek} F.\ Wilczek, Origins of mass,
Cent.\ Eur.\ J.\ Phys.\ 10 (2012) 1021--1037.

\bibitem{WBnu} A.\ Aguilar-Arevalo and W.\ Bietenholz,
  NEUTRINOS: Mysterious Particles with Fascinating Features,
  which led to the Physics Nobel Prize 2015,
  Rev.\ Cub.\ F\'{\i}s.\ 32 (2015) 127--136.
  
\bibitem{Nachtmann} O.\ Nachtmann, Elementary Particle Physics,
Springer, 1990 (Chapter 4).

\bibitem{Higgs} P.\ Higgs, Broken Symmetries and the Masses of Gauge Bosons,
  Phys.\ Rev.\ Lett.\ 13 (1964) 508--509.

\bibitem{HiggsCERN} CMS Collaboration,
  Observation of a new boson at a mass of 125~GeV with the CMS experiment
  at the LHC, Phys.\ Lett.\ B716 (2012) 30--61.\\
  ATLAS Collaboration, Observation of a New Particle in the Search for
  the Standard Model Higgs Boson with the ATLAS Detector at the LHC,
  Phys.\ Lett.\ B716 (2012) 1--29.   

\bibitem{HiggsWB} W.\ Bietenholz, The Higgs Particle: what is it, and why
  did it lead to a Nobel Prize in Physics?
  Rev.\ Cub.\ F\'{\i}s.\ 30 (2013) 109--112.
  
\bibitem{Maggiore} M.\ Maggiore,
  A Modern Introduction to Quantum Field Theory,
  Oxford University Press, 2005 (Chapter 6).
  
\bibitem{Zee} A.\ Zee, Quantum Field Theory in a Nutshell,
  Princeton University Press, 2010 (Chapter 1).

\bibitem{vdWaerden} B.\ L.\ Van der Waerden,
  Die Astronomie der Pythagoreer, North-Holland
  Publishing Company, 1951. 

\bibitem{Einstein17} A.\ Einstein, Kosmologische Betrachtungen zur
  allgemeinen Relativit\"{a}tstheorie, Sitzungsberichte der
  K\"{o}niglich Preu\ss ischen Akademie der Wissenschaften,
  Berlin (1917) 142--152.\\
  In a letter to his friend Paul Ehrenfest (who worked in Leiden,
  Netherlands), dated February 4, 1917, Einstein wrote: ``Ich habe auch
  wieder etwas verbrochen in der
  Gravitationstheorie, was mich ein wenig in Gefahr setzt, in einem
  Tollhaus interniert zu werden. Hoffentlich habt Ihr keins in Leiden,
  dass ich Euch ungef\"{a}hrdet wieder besuchen kann.''
  (I have again committed a crime in the theory of gravitation,
  which endangers me a bit of being interned to a madhouse.
  I hope you don't have any in Leiden, so I can visit you again
  without danger.)\\
  http://alberteinstein.info/vufind1/images/einstein/ear01/view/1/9396\\
  $\_$000003544.pdf

\bibitem{Friedmann} A.\ Friedmann, \"{U}ber die Kr\"{u}mmung des Raumes,
  Z.\ Phys.\ 10 (1922) 377--386; 
  \"{U}ber die M\"{o}glichkeit einer Welt mit konstanter negativer
  Kr\"{u}mmung des Raumes, Z.\ Phys.\ 21 (1924) 326--332.\\
  G.\ Lema\^{\i}tre, Un Univers homog\`{e}ne de masse constante et de
  rayon croissant rendant compte de la vitesse radiale des n\'{e}buleuses
  extra-galactiques, Annales de la Soci\'{e}t\'{e} Scientifique de Bruxelles,
  A47 (1927) 49--59.
  
\bibitem{Nuss} H.\ Nussbaumer and L.\ Bieri, Discovering the
  Expanding Universe, Cambridge University Press, 2009.
  
\bibitem{accUni} A.\ Riess {\it et al.},
  Observational Evidence from Supernovae for an Accelerating Universe
  and a Cosmological Constant, Astron.\ J.\ 116 (1998) 1009--1038. \\
  S.\ Perlmutter {\it et al.},
  Measurement of $\Omega$ and $\Lambda$ from 42 High-Redshift Supernovae,
  Astrophys.\ J.\ 517 (1999) 565--586.
  
\bibitem{Sola} J.\ Sol\`{a}, Cosmological constant and vacuum energy:
  old and new ideas, J.\ Phys.\ Conf.\ Ser.\ 453 (2013) 012015.

\bibitem{Lamoreaux} S.\ K.\ Lamoreaux,
Demonstration of the Casimir Force in the 0.6 to 6 $\mu$m Range,
Phys.\ Rev.\ Lett.\ 78 (1997) 5--8.\\
U.\ Mohideen and A.\ Roy,
Precision Measurement of the Casimir Force from 0.1 to 0.9~$\mu$m,
Phys.\ Rev.\ Lett.\ 81 (1998) 4549.

\bibitem{Casimirnovac} J.\ Schwinger,
  Casimir effect in source theory, Lett. Math. Phys. 1 (1975) 43--47.\\
J.\ Schwinger, J.\ DeRaad and K.\ A.\ Milton,
Casimir Effect in Dielectrics, Ann.\ Phys.\ (N.Y.) 115 (1978) 1--23.\\
R.\ L.\ Jaffe, The Casimir Effect and the Quantum Vacuum,
Phys.\ Rev.\ D 72 (2005) 021301.

\bibitem{Arkani} N.\ Arkani-Hamed, L.\ J.\ Hall, C.\ F.\ Kolda and
  H.\ Murayama, A New Perspective on Cosmic Coincidence Problems,
  Phys.\ Rev.\ Lett.\ 85 (2000) 4434--4437.
  
\bibitem{Carroll} S.\ M.\ Carroll, The Cosmological Constant,
Living Rev.\ Rel.\ 4:1 (2001) 1--56.

\bibitem{Witten} E.\ Witten, The Cosmological Constant from the
  Viewpoint of String Theory, Lecture at DM2000, arXiv:hep-ph/0002297.
  
\end{thebibliography}
\end{document}